\documentclass[namedreferences]{solarphysics}

\usepackage[hyperref,optionalrh,natbib]{spr-sola-addons} 
\usepackage{graphicx}        
\usepackage{color}           
\usepackage{lineno}

\newcommand{\etal}{{\it et al.}}

\begin{document}

\begin{article}

\begin{opening}

\title{Understanding the Fe {\sc i} Line Measurements Returned by the Helioseismic and Magnetic Imager (HMI)}

\author{D.~P.~\surname{Cohen}$^{1,2}$\sep
        S.~\surname{Criscuoli}$^{1}$\sep
        L.~\surname{Farris}$^{1,3}$\sep
        A.~\surname{Tritschler}$^{1}$ \sep   
       }
\runningauthor{Cohen \etal}
\runningtitle{Understanding HMI Fe {\sc i} Line Measurements}

   \institute{$^{1}$ National Solar Observatory Sacramento Peak, P.O.Box 62, Sunspot, NM, 88349-0062, USA\\
                     email: \url{danielcohen@ucla.edu} \\ 
              $^{2}$ Department of Physics and Astronomy, UCLA, Los Angeles, CA 90095-1547\\
              $^{3}$ Dept. Astronomy, New Mexico State University, P.O. Box 30001, 4500 Las Cruces, USA
             }

\begin{abstract}
The {\it Helioseismic and Magnetic Imager} (HMI) aboard the {\it Solar Dynamics
Observatory} (SDO) observes the Sun at the Fe {\sc i} 6173  {\AA} line and
returns full-disk maps of line-of-sight (LOS) observables including
the magnetic flux density, velocities, Fe I line width, line depth, and continuum
intensity. These data are estimated through an algorithm (the MDI-like algorithm, hereafter), which combines observables obtained
at six wavelength positions within the Fe {\sc i} 6173  {\AA} line.
To properly interpret such data it is important to
understand any effects
of the instrument and the pipeline that generates
these data products. 
We tested the accuracy of the line width, line depth, and
continuum intensity returned by the MDI-like algorithm 
using various one-dimensional (1D) atmosphere models.  
It was found that HMI estimates of  these quantities are highly dependent on the shape of the line,
therefore on the LOS angle and the magnetic
flux density associated with the model, and less to line shifts with respect to the central positions of the
 instrument transmission profiles.
In general, the relative difference between synthesized values 
and HMI
estimates increases toward the limb and with the increase of the
field; the MDI-like algorithm seems to fail in regions with fields
larger than approximately 2000 G.
  Instrumental effects were investigated by the analysis of HMI data obtained at  daily
intervals for a span of three years at disk center in the quiet Sun and
hourly intervals for a span of 200 hours. The analysis revealed periodicities induced by the variation of the orbital velocity of the observatory with respect 
to the Sun, and long-term trends attributed to instrument adjustments, re-calibrations and instrumental degradation.

\end{abstract}
\end{opening}


\section{Introduction}\label{S-Introduction}

The {\it Helioseismic and Magnetic Imager} (HMI) aboard the {\it Solar Dynamics Observatory} (SDO) samples 
the magnetically sensitive photospheric Fe {\sc i} 6173.34~{\AA} absorption line,
through six narrow-band filters  in right- and left- circular
  polarized light. Like the data-reduction pipeline of the {\it
    Michelson Doppler Interferometer} (MDI) on board the {\it Solar
    and Heliospheric Observatory} (SOHO) 
(which sampled the Ni 6768~{\AA}  line at five wavelength positions), the HMI-pipeline employs 
an algorithm (the MDI-like algorithm) that combines the filtergrams to obtain measurements of 
line-of-sight (LOS, hereafter) Doppler velocity, magnetic flux density,  and Fe {\sc i}
continuum intensity. 
As a by-product, the MDI-like algorithm also provides estimates of the
Fe {\sc i} line depth and width. We refer to the continuum intensity, line
depth and width as line parameters, hereafter.

The MDI-like algorithm assumes that the profile of the observed line
is Gaussian shaped. Such an assumption, together with uncertainties
in the transmission profiles of the instrument, saturation 
of the line in the presence of strong magnetic fields, and  Doppler shifts induced by 
plasma motion and solar rotation,  
are known to have caused uncertainties in measurements returned by the MDI. For instance,  
\cite{wachter2006}, \cite{rajaguru2007} and \cite{wachter2008} investigated uncertainties in Doppler velocity estimates.
\cite{tran2005}, \cite{demidov2009} and \cite{ulrich2009} investigated uncertainties in magnetic flux measurements, 
while \cite{mathew2007} and \cite{criscuoli2011} focused on uncertainties in continuum intensities. 

Similarly, recent studies have investigated the accuracy of HMI measurements. \cite{fleck2011} 
employed three-dimensional hydrodynamic simulations to study the effects of uncertainties 
in the shape and position of the HMI transmission profiles on Doppler velocity estimates. 
\cite{liu2012}, \cite{pietarila2013} and \cite{riley2014} compared HMI magnetic flux measurements 
with those obtained with MDI and other instruments and derived conversion factors between the various magnetograms. 
The analysis by \cite{liu2012} showed in particular that HMI
 magnetic flux density measurements 
are affected by 24 and 12-hour periodicities, induced by the
SDO orbital motion.

\cite{couvidat2012} employed high spatial resolution spectro-polarimetric observations of an 
active region at disk center  obtained at the {\it Dunn Solar
  Telescope} with the {\it Interferometric Bidimensional
  Spectropolarimeter} (IBIS) 
instrument \citep{cavallini2006} to investigate the effects of line shape and  
velocity on all the data products (LOS velocity, magnetic flux density and Fe {\sc i} line parameters) 
returned by the MDI-like algorithm. These authors found that the differences between 
measurements derived from the IBIS observations and those derived applying the MDI-like 
algorithm to the observed spectral profiles are in general below 20\%, 
with the best estimates obtained for the continuum intensity (below 2\%). 

Besides uncertainties resulting from assumptions in the derivation of the MDI-like algorithm, the HMI measurements 
are known to be affected by instrumental effects. These include a long-term increase
of the opacity of the entrance window, variations of the focus, and uncertainties in the 
shape and position of the filter transmission profiles (Couvidat (2014),
private communication). These effects are partially compensated for 
in the case of velocity and magnetic field measurements along with
continuum intensity measurements, but not so for the Fe {\sc i} line depth
and width. 
Moreover, the effects of variations of instrument characteristics on the HMI data product, especially on long-temporal scales 
(months to years) have not been presented in the literature yet.\footnote{Some information on instrument characteristics/calibrations
  can be found at http://jsoc.stanford.edu/}

It is important to stress that the main purpose of the HMI-pipeline is to provide Dopplergrams, magnetograms, and Fe {\sc i} continuum intensity,
while the Fe {\sc i} line width and depth 
are a by-product 
of the MDI-like algorithm. For this reason little effort 
has been dedicated to the improvement of such measurements (see also Section \ref{S-HMIpipeline}). 
On the other hand, investigations of properties of line profiles are interesting for several types of studies,
especially those in the framework of sun-as-a-star and long-term 
variations of solar magnetism 
\citep[{\it e.g.},][]{criscuoli2013, bertello2012,
  pietarila2011,livingston2007,penza2006}.

In this study we therefore investigate the accuracy of HMI data products, focusing on the Fe {\sc i} line-shape parameter estimates. To this end we employ 
synthetic Fe {\sc i} line profiles to test the MDI-like algorithm, extending the work by \cite{couvidat2012} to a larger sample of different LOS values, 
Doppler shift and magnetic field strength. 
We also employ HMI data to investigate the effects of SDO's orbital motion and 
instrumental degradation on short (days) and 
long (months to years) term measurements of the Fe {\sc i} line parameters.

The article is organized as follows: the implementation 
of the MDI-like algorithm is briefly described 
in Section \ref{S-HMIpipeline}, tests of this algorithm using synthesized line profiles are discussed
in Section \ref{S-HMIpipelinetests}, the analysis of HMI data is
presented in Section \ref{S-HMIdatatests}, and our findings and 
their implications are summarized in Section \ref{S-Conclusion}.


\section{The MDI-like Algorithm}\label{S-HMIpipeline}      

The HMI uses a narrow band tunable filter
to sample the Fe {\sc i} 6173 {\AA} line at six
different wavelength positions.

\begin{figure}    
   \centerline{\includegraphics[width=8cm]{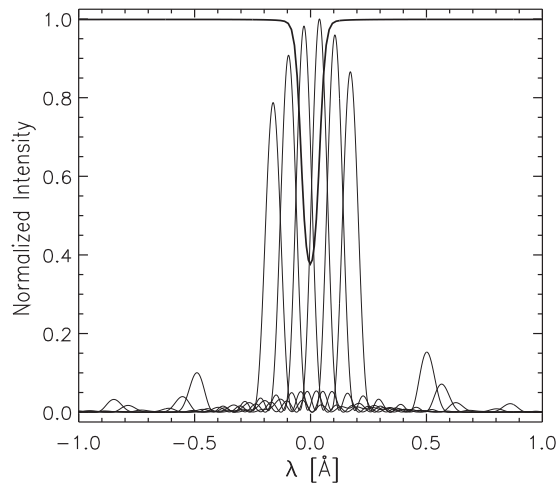}
              }
   \caption{Example of the six HMI transmission filter profiles. The thicker line is a synthetic Fe {\sc i} 6173 {\AA} line computed using a FAL-C (quiet Sun) model.}
    \label{HMItransms}
  \end{figure}

Figure~\ref{HMItransms} displays an example of the transmission profiles for each of the wavelength positions along with a synthesized iron line using a
1D quiet-sun atmosphere model (FAL-C) \citep{fontenla1999}.
 As the line is sampled in both, left and right-circular polarized
 light (LCP and RCP, hereafter), 12 filtergrams are
obtained during the wavelength tuning process. 
  After standard data correction (flat field, dark current, cosmic
  rays), these filtergrams are subsequently used to compute derived
  data products ({\it i.e.} observables)
like Dopplergrams, magnetograms, line-width, line-depth and continuum-intensity maps employing a 
MDI-like algorithm. In synthesis, the algorithm estimates the data products as functions of   
the first two discrete Fourier coefficients of the Fe {\sc i} line profile, which is assumed to be Gaussian-shaped. The Fourier coefficients are estimated by 
a proper combination of the filtergram intensities. A complete description of the algorithm and of its implementation is given in 
\cite{couvidat2012}. In the Appendix we summarize only the steps necessary for the estimation of the Fe {\sc i} line parameters. 

We implemented the algorithm in the Interactive Data Language (IDL). The main difference between our implementation and the one described in \cite{couvidat2012}
is in the HMI map employed to estimate the  $\sigma$ parameters used to compute the continuum intensity and line depth,
as explained in the Appendix. 

Estimates derived by the MDI-like algorithm are affected by uncertainties induced by the physical and numerical assumptions on which the algorithm was derived. 
Namely, these uncertainties are generated by deviations of the Fe {\sc i}
line from a Gaussian profile, the deviation of the transmission
profiles from $\delta$-functions,  the assumption that a discrete
Fourier transform can be computed based on only six data points, and the use of only two orders in the Fourier expansion. Moreover, changes of the shape of the transmission filter profiles, together with relative
shifts of the transmission filter positions with respect to the center of the line (induced either by drifts of the transmission profiles or Doppler effects),
also affects the estimates of the data products. The HMI-pipeline employs proper look-up tables to partially compensate for such effects. Nevertheless, these corrections
are applied only to the Doppler velocity and magnetic field and thus partially to the
  Fe {\sc i} continuum intensity, which is function of the Doppler
  shift (see Equation (\ref{eq-continuum}) in the Appendix). 
The Fe {\sc i} line depth and width estimates are
a by-product of the MDI-like algorithm and no further correction is applied to them.

\section{Tests on Synthetic Profiles} \label{S-HMIpipelinetests}

In order to investigate the effects of the deviation of the line shape
from a Gaussian and the effects caused by the shift of the line with respect to the 
HMI wavelength filter positions we used synthesized profiles of the Fe {\sc i} 6173 {\AA}  line as an input to the 
MDI-like algorithm.
The synthesized input profiles were computed in non-local
thermodynamic equilibrium with the RH code 
 \citep{uitenbroek2001} on a 3 {\AA} wide spectral range centered around the line.

\begin{table}

	\centerline{}
	\begin{tabular}{c c c}
	\hline\hline
	Model & Description & $B$ [kG] \\
	\hline
	FAL-A & quiet Sun & --- \\
	FAL-C & quiet Sun & --- \\
	FAL-E & faint network & --- \\
	FAL-P & facula & 1.00 \\
	Kur\_5500 & sunspot & 1.10 \\
	Kur\_5250 & sunspot & 1.70 \\
	Kur\_4750 & sunspot & 2.25 \\
	Kur\_4500 & sunspot & 2.50 \\
	Kur\_4250 & sunspot & 3.00 \\
	Kur\_4000 & sunspot & 3.50 \\
	Kur\_3750 & sunspot & 4.00 \\
	\hline
	\end{tabular}
	\caption{1D atmosphere models used to synthesize the Fe {\sc i}
          line and corresponding vertical magnetic field values \citep{fontenla1999}.
	The Kurucz models are denoted Kur\_XXXX where XXXX is effective temperature in Kelvin. }
	\label{tab_Models}
\end{table} 

\begin{figure}    
   \centerline{\includegraphics[width=5cm, trim=0.9cm 0cm 1.0cm 0.5cm]{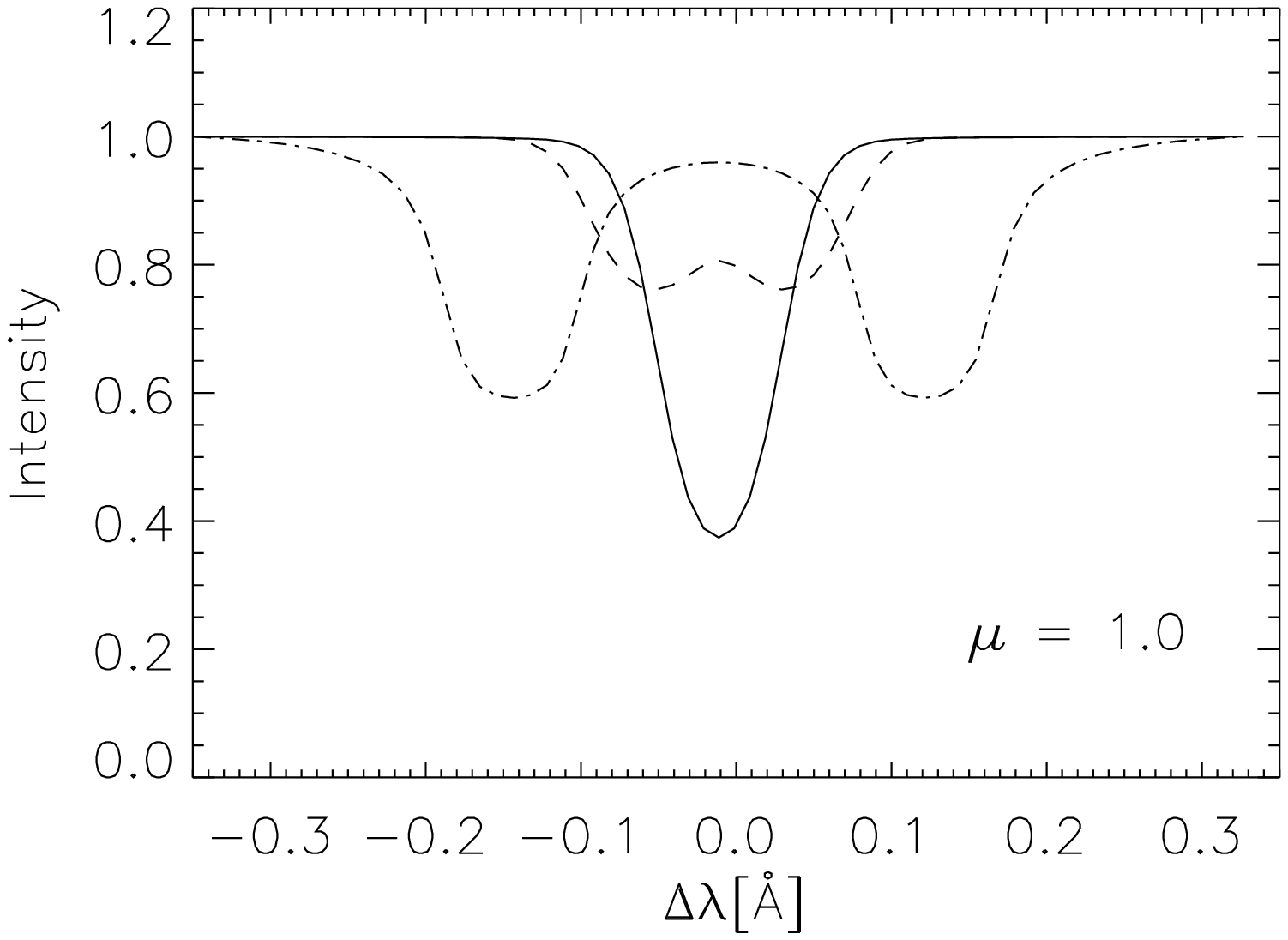}\includegraphics[width=5cm,trim=0.9cm 0cm 1.0cm 0.5cm]{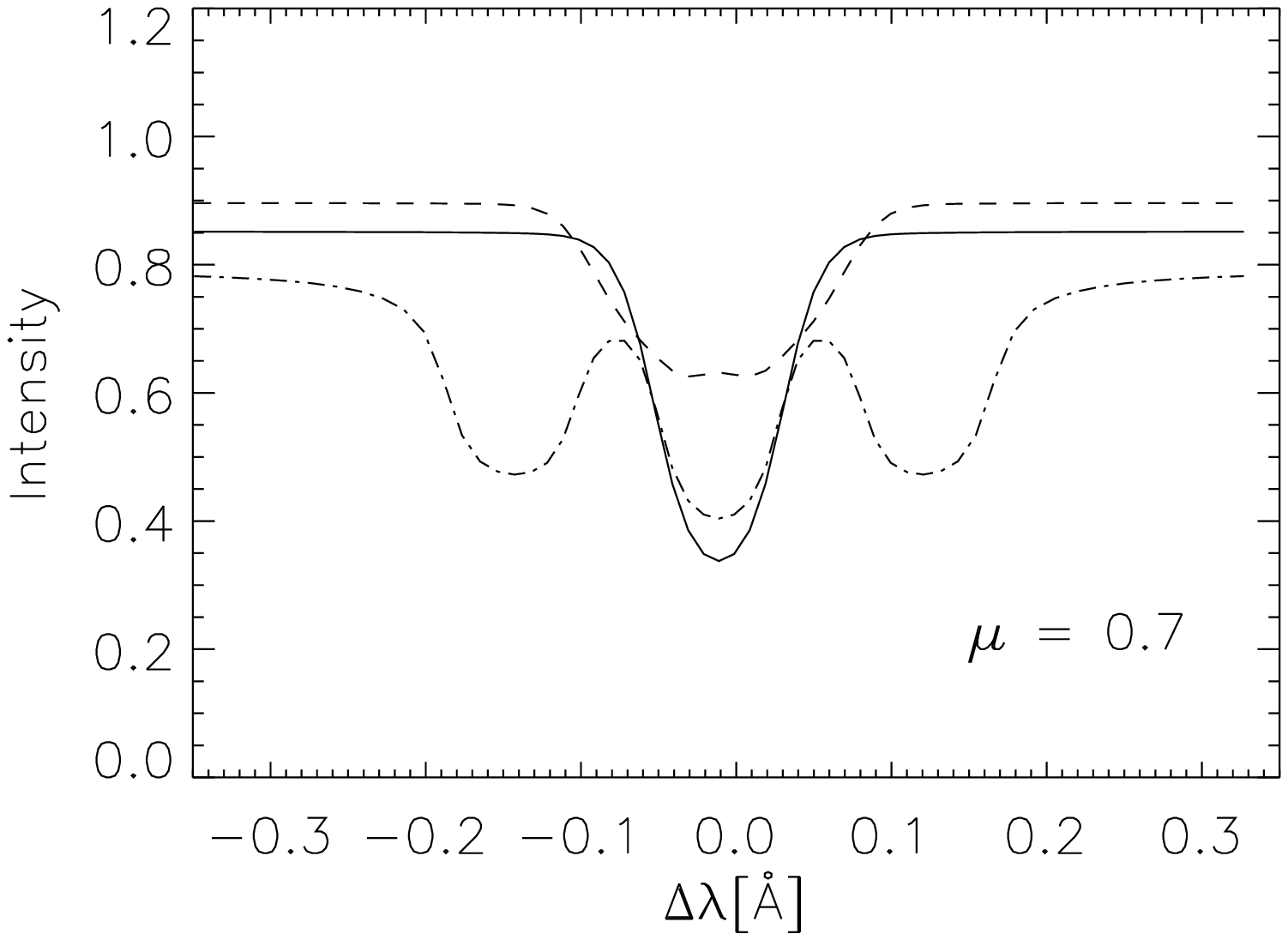}
   \includegraphics[width=5cm,trim=0.9cm 0cm 1.0cm 0.5cm]{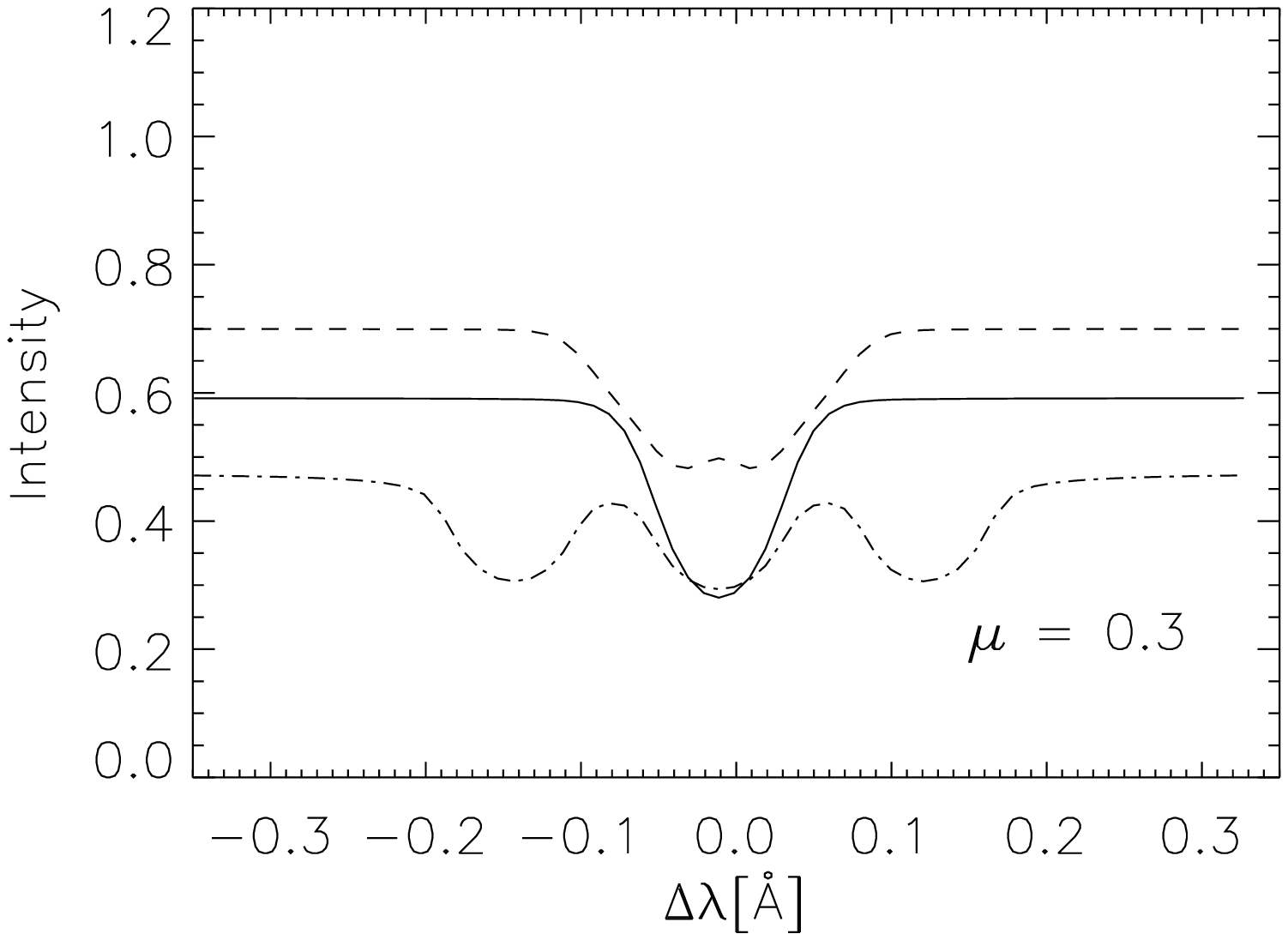}
              }
   \caption{Examples of Fe {\sc i} line profiles synthesized through
     three 1D atmosphere models at three viewing angles. Continuous
     line: FAL-C. Dashed line: FAL-P with
   uniform magnetic field of 1000 G.
   Dot-dashed line: Kurucz\_4250 with uniform magnetic field of 3000 G. All profiles have been normalized to the corresponding continuum intensity at disk center.}
    \label{Lineshapes}
 \end{figure}

Since the shape of a spectral line is determined by the physical properties of the plasma, and, 
in the case of magnetically sensitive lines, by the magnetic field strength
we synthesized the line using 
various static one-dimensional atmosphere models 
representing different features observed on the solar disk.
Quiet sun, network and facular region profiles were synthesized using FAL99 
models \citep{fontenla1999}, while sunspot profiles were synthesized using Kurucz 
models characterized by various effective temperatures. We also performed 
tests on the three Maltby models as described in  \cite{criscuoli2011}, nevertheless 
results obtained with these latter models are qualitatively in 
agreement with those found from the Kurucz models and therefore they are not discussed here. 
To investigate the effects of the Zeeman splitting on lines, the facular model FAL-P and the sunspot 
models were synthesized imposing a vertical magnetic field of constant
strength. Although the imposed magnetic fields are arbitrarily
associated with the model effective temperature, the values do follow
the widely known anti-correlation between magnetic field and
temperature in sunspots \citep[{\it e.g.},][and references therein]{schad2014}.
The models employed and the corresponding magnetic field strengths
are summarized in Table \ref{tab_Models}. 
For each model we considered eight viewing angles, 
expressed in 
the following as the cosine of the heliocentric angle, $\mu$. 
  Note that by synthesizing the line for different viewing angles we
  equivalently investigate the effect of inclined magnetic fields. Examples 
of line profiles generated using these models are shown in 
Figure \ref{Lineshapes}. 
To investigate 
the effects of velocities due to convective motions, 
solar rotation and relative motions between the observatory and the Sun,
we also shifted the line profiles 
between [-4,4] km s$^{-1}$ (note that the SDO velocity relative to the Sun spans 
between -3.2, and 3.2 km s$^{-1}$). The line parameters were estimated by feeding the synthetic profiles to the MDI-like algorithm and the obtained values
were compared with those derived from direct estimates on the synthetic profiles (synthetic values, hereafter). 
In particular, following the definitions of the MDI-like algorithm
(see Appendix) for the line parameters, we defined the synthetic
continuum as the intensity synthesized  at 0.5 {\AA}
from line center, the line depth as the difference between the emergent intensities in the continuum and at the center of the line, 
while the line width was estimated as the full width at half
maximum. All line parameters were derived separately for the LCP and
RCP components of the spectral line and then averaged to give the
final value.

\subsection{Line Shape Effects: Model and Line-of-Sight Dependence}\label{sec:los}

\begin{figure}    
   \centerline{\includegraphics[width=12cm]{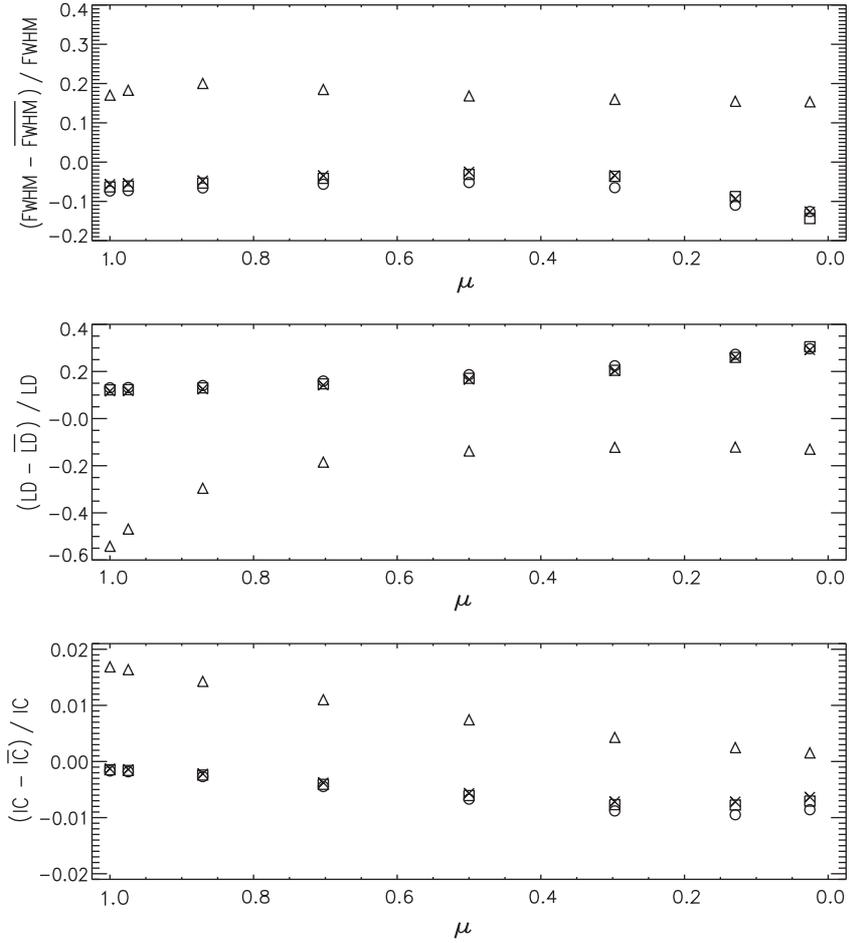}
              }
   \caption{Relative differences between synthetic and MDI-like algorithm-calculated line parameter values for the FAL99 models: FAL-A (open circle), FAL-C (open square), FAL-E (X), and FAL-P 
   (open triangle). FWHM denotes the line-width, LD the line-depth and
   IC the continuum intensity. The bar on top of the acronyms denotes quantities returned by the MDI-like algorithm.}
    \label{FALObs}
  \end{figure}

Figure \ref{FALObs}  shows the relative 
difference amongst  the line parameters derived with the MDI-like algorithm and the synthetic ones, 
for the FAL99 models and for various lines-of-sight.
The plots show that the relative difference obtained 
from FAL-A, FAL-C, and FAL-E are similar and that in particular 
the MDI-like algorithm on these 
three models overestimates the line width by 5--10\% and 
the line depth by 10--30\%, with a higher relative difference in the line depth
resulting at shallower lines-of-sight (smaller $\mu$ values).
Continuum intensity is found to be more accurate as the difference 
between synthetic values and  MDI-like algorithm estimates are within 1\% 
with a minimum of about 0.1\% close to disk center.
The differences obtained with the FAL-P model representative of a 
facular region with a vertical magnetic 
field of 1000 G, are larger and have opposite sign with respect 
to the other FAL models investigated. 
This is because the Fe {\sc i} line shape profiles 
obtained from the first three models are similar 
and close to the shape of a Gaussian profile
whereas the shape obtained from the FAL-P is broadened by the 
Zeeman effect (as shown in Figure \ref{Lineshapes}). 
The line width is in fact underestimated by 15--20\%, the line depth 
is overestimated by 20--60\%, and the continuum intensity is
underestimated by 0.5--2\%. The differences for this model decrease 
toward the limb, because, as shown in Figure \ref{Lineshapes}, the line core intensity enhancement due to the Zeeman splitting decreases for shallower LOS.

    \begin{figure}    
   \centerline{\includegraphics[width=12cm]{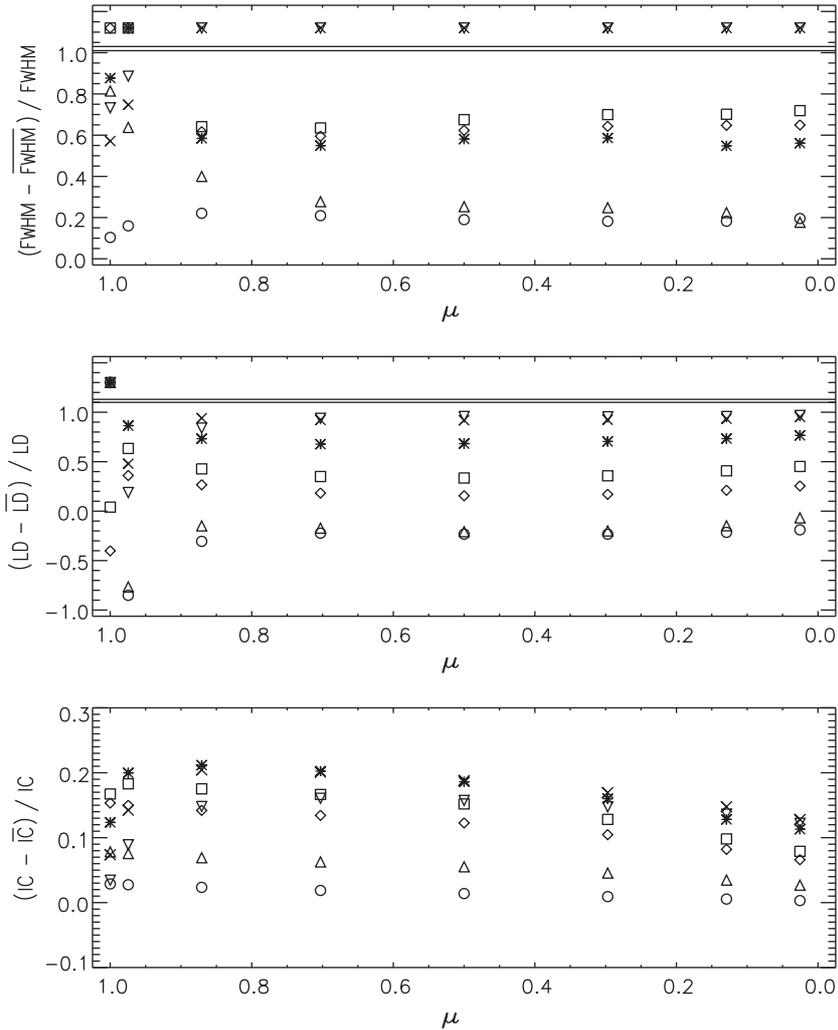}
              }
   \caption{Relative differences between theoretical and MDI-like algorithm-calculated line shape values for the Kurucz sunspot models: Kur\_5500 (open circle), 
   Kur\_5250 (open upward triangle), Kur\_4750 (open diamond), Kur\_4500 (open square), Kur\_4250 (star), Kur\_4000 (‘X’), Kur\_3750 (open downward triangle). 
   FWHM denotes the line-width, LD the line-depth and IC the continuum
   intensity. The bar on top of the acronyms denotes quantities returned by the MDI-like algorithm.}
    \label{KURObs}
  \end{figure}

The results obtained for all seven Kurucz models are plotted 
in Figure \ref{KURObs}. The relative difference
between the synthetic
values and those returned
by the MDI-like algorithm are very large, with the lowest values 
found in the case of the continuum intensity, for which 
the discrepancy is up to 20\%. 
For the line depth 
and the line width the differences are up to 100\%. In general, 
the differences increase with the increase of the magnetic field strength, as a consequence of both, the line saturation and the poor sampling
of the wings and continuum in the case of large line broadening. 
In particular, for magnetic field intensities larger than 
approximately 2000 G the MDI-like algorithm returned unphysical 
line width estimates. 
These cases are
marked above the double line in the plots. This occurs because 
for these models the sum of the squares of the second Fourier 
coefficients is greater than the sum of the squares of the first 
Fourier coefficients, so that the argument of the logarithm in Equation (\ref{eq-sigma}) in the Appendix
returns a negative number whose square root is not real. Similarly, the 'infinite' relative difference found at vertical lines-of-sight for the line depth (these cases are
also beyond the double line in the plot) are due to the saturation of
the line at large magnetic field strengths (see Figure \ref{Lineshapes}), so 
that the synthetic line depth tends to zero, 
while the MDI-like algorithm, expecting a Gaussian-shaped profile, returns a larger value. It is also 
interesting to note that for the largest 
magnetic field strengths investigated the 
deviations from the synthetic values of the line depth  and of the line width present small center-to-limb variation because of saturation effects. 
For the same reason, as already pointed out in previous
studies investigating  the accuracy of the MDI algorithm 
\citep{mathew2007,criscuoli2011}, the discrepancies of the 
continuum intensities do not increase monotonically
with the increase of the magnetic field strength, instead they 
decrease for the largest field values investigated.

\subsection{Line-Shift Effects}\label{sec:doppl}

\begin{figure}    
   \centerline{\includegraphics[width=10cm]{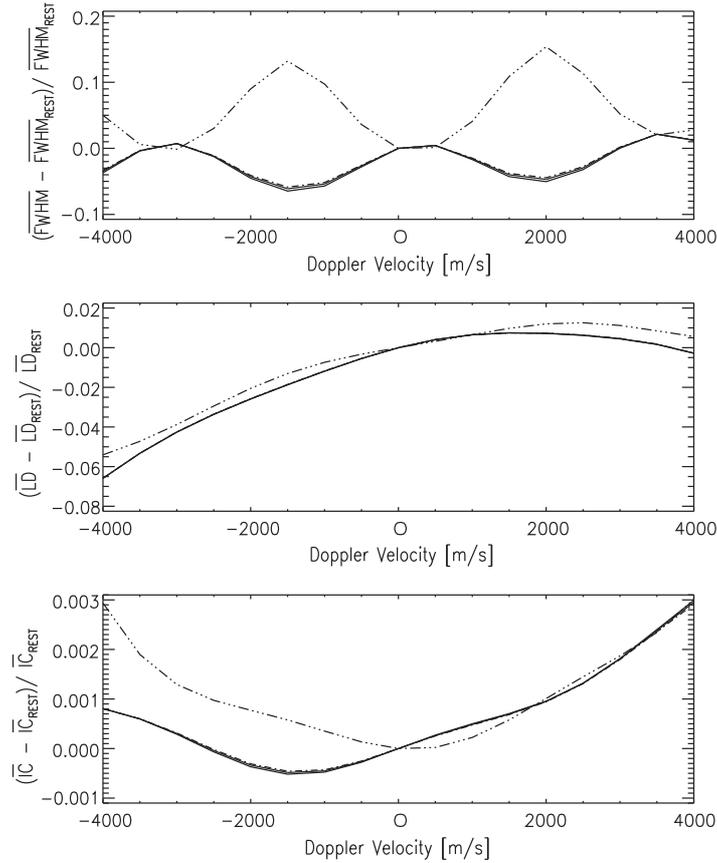}
              }
   \caption{Relative differences of line parameters  calculated with
     the MDI-like algorithm between Doppler shifted line and line at rest, for the Fe {\sc i} line generated 
   by the Fontenla models at $\mu$ = 1. Solid line: FAL-A. Dashed
   line: FAL-C. Dot--dashed line: FAL-E. Triple-dot--dashed
   line: FAL-P.}
    \label{velofal}
  \end{figure}

 We also investigate the influence of a line shift on the determination of the 
line parameters. In the following we report only results obtained with the FAL99 models, 
as those obtained for sunspots are qualitatively in agreement.
Figure \ref{velofal} shows the relative difference between the values obtained from the 
MDI-like algorithm with the line at rest and the line shifted
{\it versus} the amount of shift.
The plots reveal a periodicity 
for all the line parameters, which 
 is proportional to the spacing between the HMI transmission profiles \citep[see also][]{wachter2008,criscuoli2011},.

The variations obtained 
for models FAL-A, -C, and -E are similar in amplitude (a few percent) and sign, 
whereas the results obtained with the FAL-P model show larger deviations (up to 13\%) and opposite sign, in agreement with the results presented in Figure \ref{FALObs}. 
The increase of the difference between measured and synthetic values is due in this case to the fact that a line shift causes the left and right wings
of the line to be unevenly sampled. The broadening induced by the Zeeman splitting enhances this effect (for larger broadenings even modest shifts may 
cause portions of one of the wings not to be sampled at all), so that it is no surprise that the uncertainties increase with the increase of the 
magnetic field strength associated with the model.

In agreement with results presented in \citet{criscuoli2011} for the continuum intensity estimated by the MDI algorithm, 
the plots also show asymmetries around zero velocity.  \citet{criscuoli2011} 
ascribed this
to the fact that a line shift could compensate for or enhance the original shift of the lines that they employed (obtained from non-static simulations and observations)
for their tests. Because the results presented in Figure~\ref{velofal} have been obtained with static models, the asymmetries around the zero velocity must be ascribed to
small asymmetries in the filter transmission profiles.

Finally, it is worth to note that the results reported in Figure~\ref{velofal} allow to address 
effects introduced by  line shifts only.
In practice, uncertainties are introduced by the combined effect of line-shape and line-shift. For each model and shift, it is easy to show that,
if $h$ is the uncertainty  resulting from the deviation of the line shape from a Gaussian profile  (results reported in Figure~\ref{FALObs}), 
and  $h'$ is the uncertainty resulting
from a line-shift (results reported in Figure~\ref{velofal}), then the uncertainty resulting by the combination of the two effects is $h + h' +h\cdot h'$.

\section{Tests on HMI Data} \label{S-HMIdatatests}

The results presented in the previous section indicate that the accuracy with which the  MDI-like algorithm estimates the Fe {\sc i} line parameters 
strongly depends on the shape of the line profiles and the line shift. The first effect dominates, at 
least for shifts in the range investigated. We expect studies aiming at investigating the center-to-limb variation or
the temporal evolution of magnetic features to be particularly affected.
For instance \cite{liu2012} showed that the evolution of the LOS
magnetic flux density of an active region (AR) measured by HMI presents 
12 and 24 hours cyclic variations,
which are generated by the orbital motion of the SDO. This signal is super-imposed to a longer temporal trend  
resulting by the combination of the change of the LOS with solar rotation Doppler effects  as the region crosses the solar disk. 
To investigate these effects also on
the line-shape parameters returned by the HMI, we repeated the analysis performed by \cite{liu2012} on a
different region ({\it i.e.} AR 11092) and investigated the
evolution of the line parameters and of the magnetic flux density on 
quiet, facular, and sunspot regions.

We also analyzed 
the variation of the line-shape parameters over a data set of HMI 
observations spanning about three years of observations with the aim to investigate the effects of instrumental 
degradation described in Section \ref{S-Introduction}.  
As properties of magnetic
features can vary over the cycle, for this last analysis only 
quiet-sun regions at disk center were considered.
 
\subsection{Short-Term Line Parameters Variations}

\begin{figure}    
   \centerline{\includegraphics[width=7.5cm]{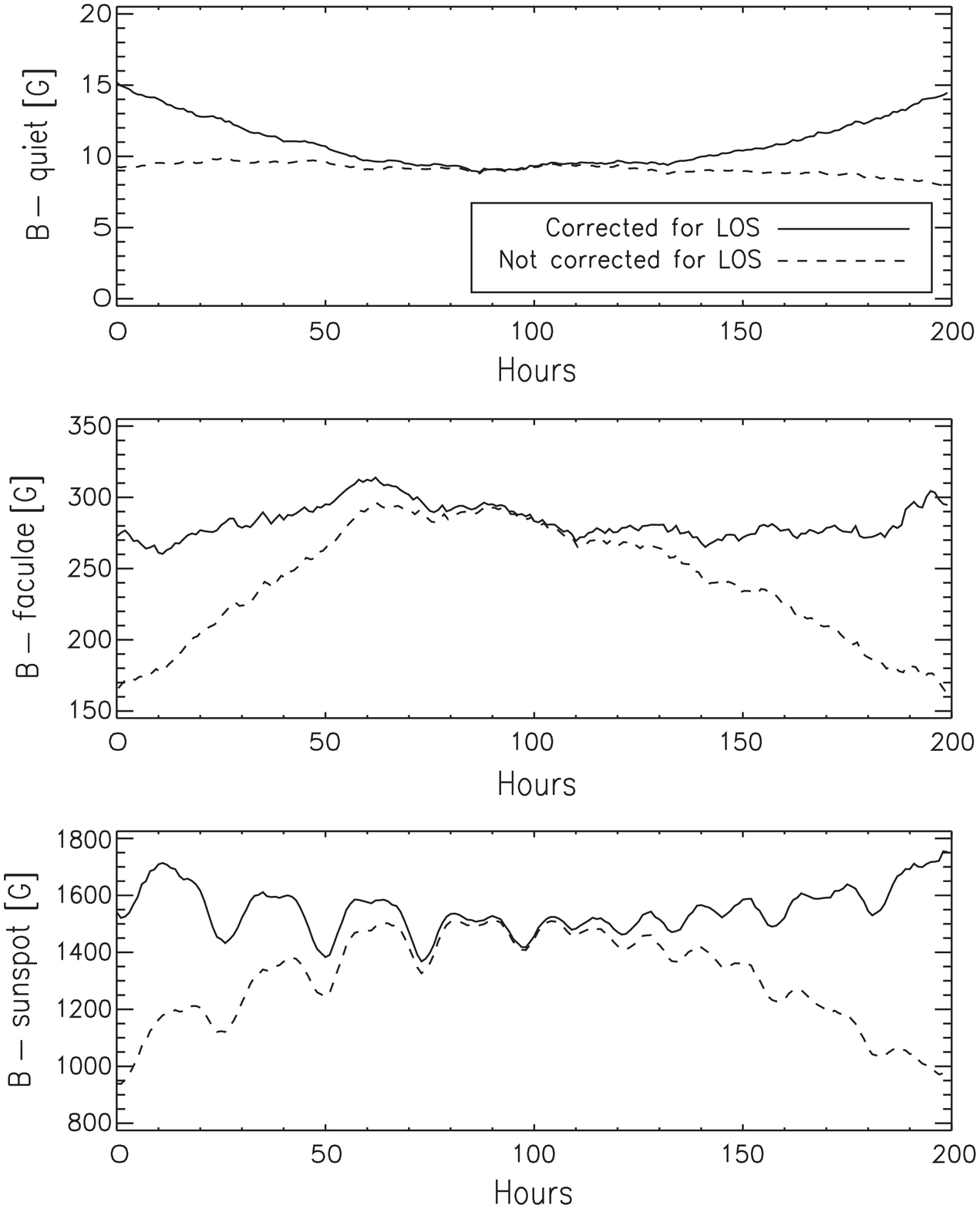}{\includegraphics[width=7.5cm]{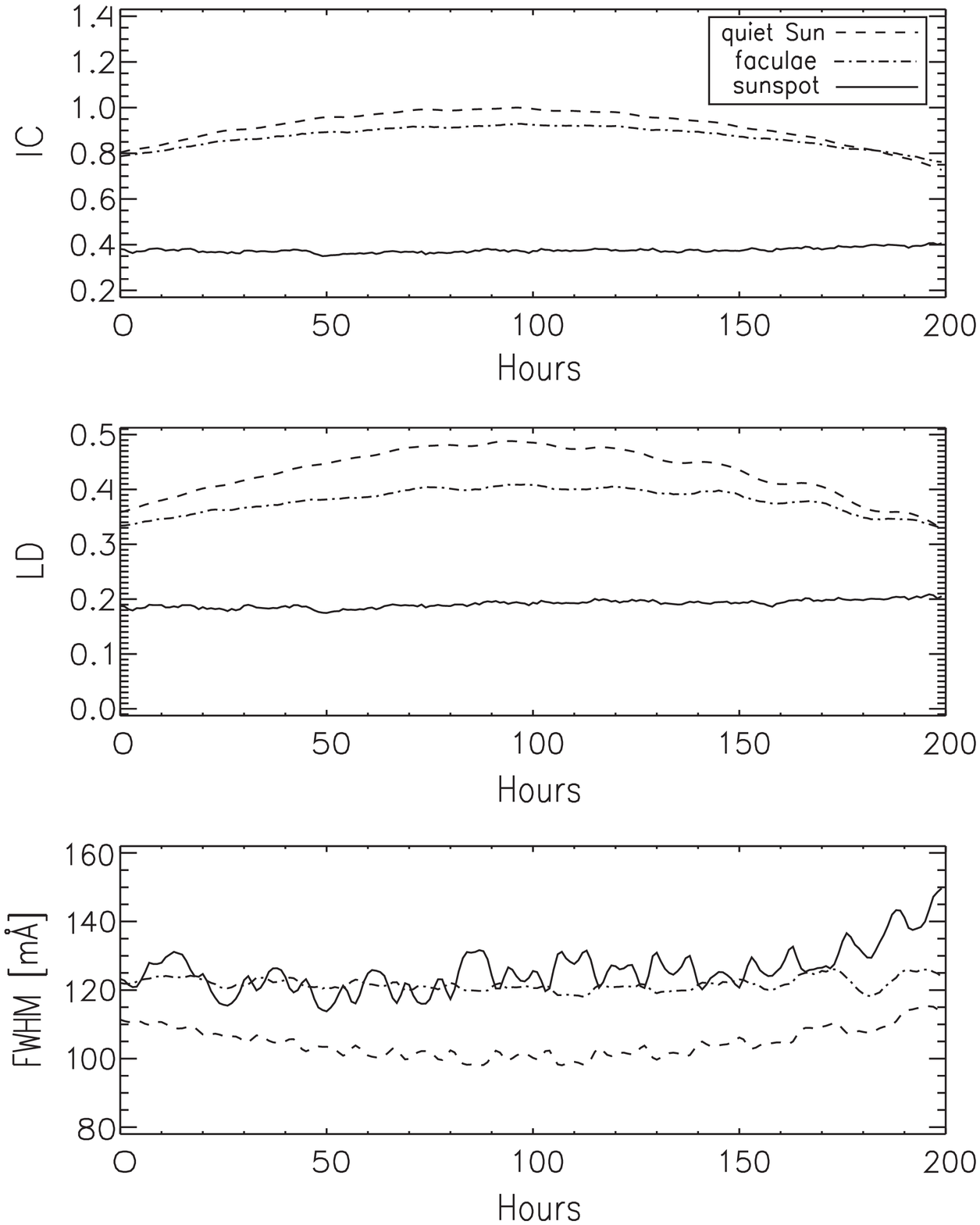}
              }
              }
   \caption{Variation of average magnetic flux density (left) and average line shape parameters (right) of quiet, facular and sunspot pixels (see text) observed around AR11092 by the HMI.
   The observations started on July 30, 2010 at 11:58:26 UT. The region was at the closest proximity  to disk center about 100 hours after the beginning of the observations. 
   Continuum intensity and line depth are normalized to the average continuum intensity of quiet pixels at disk center.}
    \label{ShortObs}
  \end{figure}

In order to investigate short-term effects of solar rotation and
orbital velocities on HMI data products we analyzed a one hour cadence  data set
spanning about 200 hours  in total, acquired between  July 30 and August 8,
2010. In particular, we analyzed the 720
sec magnetic flux, line-depth, continuum intensity and line-width
data products. We employed magnetograms corrected for the
fore-shortening ($B = B_{obs}/\mu$) to identify quiet (pixels where
absolute magnetic flux density is $\leq$  100 G), facular (pixels where
absolute magnetic flux density  is $>$ 100 and $\leq$ 1000 G) and sunspot (pixels
where absolute magnetic flux density is $>$ 1000 G)
regions around AR 11092 (NOAA number).  This active region was chosen because it showed little evolution
during its passage over the disk.  For each class of pixels we then
studied the temporal variation of the average Fe {\sc i} line parameters and the
magnetic flux density. The investigation of the average magnetic flux density allowed
verifying that the  observed line parameter variations were not
caused by the evolution of the magnetic field in the region.  
The variations of the average magnetic flux density computed over the three classes of regions are
reported in Figure \ref{ShortObs} (left panel). 
The dashed and continuous lines represent the magnetic flux density non-corrected and
corrected for the fore-shortening, respectively.  These plots agree
with results reported in \cite{liu2012}, which showed that the
magnetic flux density of an active region  non-corrected for the fore-shortening 
is maximum at disk center, while the flux density of  quiet regions does not  depend
on the position on the disk. When correcting for the fore-shortening,
the curves change concavity, with a minimum observed at disk center,
suggesting that the trends reported in \cite{liu2012} are mostly
caused by projection effects. Semi-diurnal and diurnal periodic variations of the flux are also observed. These, as explained in
\cite{liu2012},
can be mostly ascribed to spacecraft orbital
motions effects not entirely compensated for by the look-up
tables.
The mostly symmetric trends around disk
center suggest that  the magnetic field of the region stayed fairly
constant during the observation time.

The right panel in Figure \ref{ShortObs} shows results obtained for the
line parameters.  The plots show that, like the magnetic flux density, line
parameters are symmetric around the time of closest proximity to disk
center.  These variations must be mostly ascribed to the combination 
of LOS effects and solar rotation. The typical
periodicities caused by orbital motions of the satellite are also observed. 
These are more noticeable in the case of the line width in magnetized pixels, as
expected by results discussed in Section \ref{sec:doppl}.  It is
important to notice that all the observed temporal variations 
are due to both LOS and velocity effects and that it is not
possible, in practice, to distinguish between them. However,
comparison of results presented  in Section \ref{sec:los} and
Section \ref{sec:doppl} suggests that the longer temporal trends
of the continuum intensity and line depth reported in Figure \ref{ShortObs}
are mostly caused by LOS effects rather than by the solar
rotation, because variations induced by this last effect (the solar rotation is $\pm$ 2 km s$^{-1}$ at the equator)  are about one order of
magnitude smaller than variations induced by LOS effects. As a consequence, 
because, as shown in Section \ref{sec:los}, the shape of a line in high magnetized regions is dominated by the Zeeman splitting rather than by
LOS effects, the line parameters show almost no variation in sunspot
pixels.

   %

\subsection{Long-Term Line Parameters Variations}

\begin{figure}    
   \centerline{\includegraphics[width=10cm]{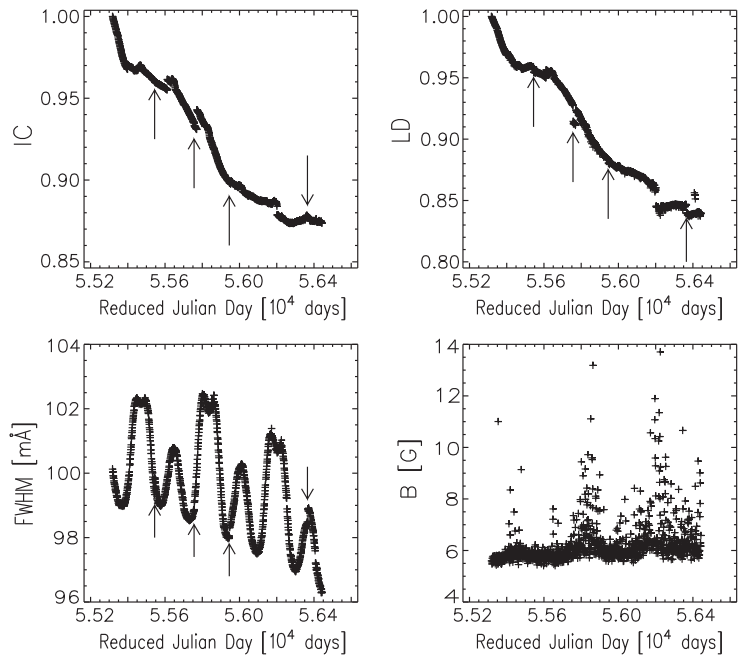}
              }
   \caption{Line parameters and average absolute magnetic flux density observed by the HMI in quiet pixels close to disk center over a period of three years.
   The arrows indicate the days in which the HMI filters were re-tuned. }
    \label{long}
  \end{figure}

In order to investigate long-term variations of the Fe {\sc i} line 
parameters returned by the HMI, we have analyzed a set of magnetic 
flux density, line-depth, continuum intensity and line-width data acquired 
daily between  May 1, 2010 and May 30, 2013. For this analysis we 
also employed 720 sec data, obtained
as closest as possible to
12:00 UT. In order to more easily  distinguish between effects 
inherent to the measurements from those induced by variations 
of magnetic feature properties induced by the solar magnetic cycle, we restricted the analysis to quiet pixels 
located close to disk center. To this end
we analyzed the temporal variation of the average value of the Fe {\sc i} data products computed 
over pixels where the absolute magnetic flux density is $\leq$ 100 G, located 
in a square area of 500$\times$500 pixels (about 250$\times$250 arcsec) 
around disk center.  The results are shown in 
Figure~\ref{long}. The line-shape parameter values 
have been normalized to the value measured the first day analyzed.  
The plots of line-shape parameters 
show an overall decrease with time and 
several discontinuities.  The long-term trends are caused by various
issues, mentioned in Section \ref{S-Introduction}, that 
are known to affect the HMI, such as an increase 
of the opacity of the entrance window (most likely induced by  
exposure to UV radiation), changes in the shape and positions of the transmission 
profiles 
and variations of the focus caused by temperature changes in the 
entrance window especially after eclipse periods.  
The discontinuities are due to periodic re-calibration of the instrument 
that are meant to mitigate these effects. The re-calibrations include:  an increase
 of the standard exposure time, retuning of the transmission
 profiles (the days of the retuning are  marked with arrows in the
 figure), adjustments of the focus, and front window 
 re-heating after eclipse periods (calibrations calendar and
 descriptions are available at
 http://jsoc.stanford.edu/doc/data/hmi/). 
Among these effects, those
 that dominate the long-term trends are the increase of the entrance
 window opacity and the shift  of the transmission profiles. 
Comparison of plots in Figure~\ref{long} and Figure~\ref{velofal}
indicate that the continuum intensity and line depth are mostly 
affected by the increase of the entrance window opacity rather than by a shift of the transmission profiles. The observed 
long-term decrease of these quantities is in fact about 10\%. Plots in Figure~\ref{velofal} show that in the case of the line depth 
such variation can be generated by shifts larger than about 4 km s$^{-1}$
(which seems an unlikely large value, considering that, as stated above, the transmission profiles are periodically re-tuned),
while the continuum intensity is almost insensitive to relative shifts of the line with respect 
to the positions of the transmission profiles. 
Instead, Equation (\ref{eq-sigma}) in the Appendix indicates that the line-width is insensitive to intensity variations, as long as 
these variations are the same for all the six filtergrams. The observed decrease of about 4\% for this parameter is compatible with shifts of the transmission profiles
smaller than 1 km s$^{-1}$. 

It is also noteworthy that the observed  variations of the line parameters 
cannot be ascribed to variations of the quiet-Sun magnetic field.
In fact, Figure~\ref{long} shows that the average absolute magnetic flux density slightly increases with time, which would cause both the line 
width and depth to increase. Note also that, as mentioned above, the magnetic flux density measurements are compensated for line shift effects (induced by both 
LOS orbital velocity changes and drifts of the transmission profiles) and that the magnetic flux density is independent from the absolute intensity of 
the filtergrams \cite[see][]{couvidat2012}, so that the small flux-density increase over time, in agreement with the increase of the magnetic activity during 
the period analyzed, could be real.

Finally, 
the plots in Figure~\ref{long} show periodic variations of the line 
parameters, which are particularly noticeable in 
the case of the  line width. A comparison of the Fourier spectra of 
the SDO velocities and of the line width (not shown) confirmed that 
even these oscillations are due to the relative variation of the SDO orbital velocity with respect to the Sun.

\section{Discussion and Conclusion}\label{S-Conclusion} 
 
In this study we investigated the accuracy of the Fe {\sc i} 6173 {\AA}
line-parameter values ({\it i.e.} line depth, line width, continuum intensity)
returned by the HMI. To this aim, we first tested the accuracy of estimates returned by the MDI-like algorithm, which is part of the HMI data reduction pipeline. 
We then identified systematic and instrumental effects by studying the temporal evolution of HMI data obtained at different cadence over short (days) and long (years) 
temporal ranges. 

An MDI-like
algorithm
was implemented and tested with synthetic Fe {\sc i} 6173 {\AA} line profiles
generated using 1D atmosphere models representing quiet-sun, faint
network, facular, and sunspot regions at different viewing angles across
the solar disk. 
We found that the accuracy of this algorithm strongly depends
on the line shape determined by magnetic field strength and
viewing angle. The algorithm works relatively well in quiet and facular regions,
where the accuracy of the continuum intensity is better than 2\% and
the accuracy of the line depth and line width are on average of the order
or below 20\%.  The accuracy decreases with the increase of the
magnetic field strength, so that in highly magnetized
regions ({\it e.g.} sunspots) it is in general of the order of  several tens of per
cent, with the best accuracy found again for the continuum intensity
(about or below 20\%). For a magnetic field strength larger than about
2000 G  the  algorithm
encounters numerical problems due to the saturation of the line, so
that in these regions HMI results cannot be considered reliable. 
These results are in partial agreement with those obtained by \cite{couvidat2012}, who employed observed spectra to investigated the accuracy of the 
data products returned by the MDI-like algorithm. In particular,  the accuracies that we estimated for the continuum intensity along vertical lines-of-sight
are in good agreement with those presented by those authors. In the case of line width and depth, the agreement is still
good in  quiet and facular regions, but in regions of stronger magnetic field we obtained a smaller accuracy than the one reported by \cite{couvidat2012}. The 
discrepancy between our results and those  presented by those authors has to be ascribed in part to
uncertainties in the radiative synthesis and assumptions in our models
({\it e.g.} the atomic transitions values employed, the arbitrary
association of an atmosphere model with a magnetic field strength,
the assumption of a vertical constant field, and the non-inclusion of molecules in umbra models). On the other hand, \cite{couvidat2012}
estimated a better accuracy in highly magnetized regions because they employed observed spectra on which the line shape variations induced by the Zeeman splitting 
(especially the increase of the core intensity)
are strongly reduced in amplitude by scattered 
light (both spatial and spectral) and 
finite spectral resolution effects \cite[see also the discussion in][] {criscuoli2011}.

The effects of the line shift 
with respect to the nominal wavelength position of the filter transmission profiles
have also been
investigated, and they resulted to be in general smaller than  LOS and
magnetic field effects, at least for shifts in the range [-4,4] km s$^{-1}$.

Along with testing the HMI pipeline, we studied Fe {\sc i} line width, line
depth, and continuum intensity HMI data generated at both hourly
intervals over a span of 200 hours in an active region and at daily
intervals for a span of three years in quiet regions at disk center. 
We found both short-term and long-term periodicities caused by the orbit
of the SDO, long-term decrease in intensity measurements due to
opacity of the front window increasing, long-term decrease in line
width due to drift in wavelength of transmission filters, along with
discontinuities present in the long-term data due to instrument
adjustments and re-calibrations. The LOS magnetic field  and the LOS Doppler velocity are partially corrected for such
effect, while the Fe {\sc i} line parameters data are not. This complicates the analysis of temporal variations of the iron line parameters.

Thus, the Fe {\sc i} line parameter estimates returned by the HMI are affected by uncertainties due to approximations on which the MDI-like algorithm works (mainly, 
the assumption that the line is Gaussian shaped) and to instrumental effects. These uncertainties could be reduced with the use of proper look-up tables, like those employed
to correct LOS magnetic field and velocity estimates and
thus continuum intensity measurements. However, such multi-dimensional look-up tables
would likely be quite difficult to implement.
Alternatively the algorithm might benefit from being modified
as to fit the actual Fe {\sc i} line profile rather than assume a 
Gaussian \citep{couvidat2012}.  In this case the speed of the fitting
algorithm would be a decisive measure of its utility, as it
would require fitting a very large number of pixels.
The long-term effects induced by instrumental degradation, especially those induced by the increased opacity of the entrance window,  are more
difficult to correct for. It is possible to make such corrections if it is assumed that the quiet Sun does not vary, perhaps committing a small
error, but long-term studies of the quiet Sun would therefore be
comprised. 
Because such corrections
are not available yet, results presented in this study are important
for the correct interpretation of the Fe {\sc i} 6173 {\AA}  line-parameter estimates returned by the HMI.

\begin{acks}
The autors are grateful to Sebastien Couvidat for the valuable comments on the manuscript.  
This work was
carried out through the National Solar Observatory Research
Experiences for Undergraduate (REU) site program, which is co-funded by the
Department of Defense in partnership with the NSF REU Program. The National
Solar Observatory is operated by the Association of Universities for Research in
Astronomy, Inc. (AURA) under cooperative agreement with the National Science
Foundation.
\end{acks}



\bibliographystyle{spr-mp-sola}
\tracingmacros=2
\bibliography{HMIbib}
 

\appendix
\label{append}
\section{The MDI-like algorithm}
We implemented the MDI-like algorithm following the description in
\cite{couvidat2012}. Here we summarize only the steps necessary to
estimate the Fe {\sc i} line parameters.

The algorithm assumes a Gaussian shaped Fe {\sc i} line profile:
\begin{equation} \label{eq-Gaussian}
	I(\lambda) = I_c - I_d \exp{\left[-\frac{(\lambda - \lambda_0)^2}{\sigma^2}\right]},
\end{equation}
where $I_c$ is the continuum intensity, $I_d$ is the line depth, and $\sigma$ is a parameter that defines the width. 
Note that the algorithm returns as an estimate of the line width, the full width at half maximum 
(FWHM), which  is proportional to $\sigma$ through the relation 
$FWHM = 2\sqrt{2\ln{(2)}}\sigma$. 

The algorithm first computes discrete approximations to the first and second Fourier coefficients using the six filtergram intensities $I_j$:
\begin{eqnarray} 
	a_1 \approx \frac{2}{6}\sum_{j=0}^{5} I_j \cos{\left(2\pi \frac{2.5 - j}{6}\right)}, && b_1 \approx \frac{2}{6}\sum_{j=0}^{5} I_j \sin{\left(2\pi \frac{2.5 - j}{6}\right)} \label{eq-Fourier1}\\
	a_2 \approx \frac{2}{6}\sum_{j=0}^{5} I_j \cos{\left(4\pi \frac{2.5 - j}{6}\right)}, && b_2 \approx \frac{4}{6}\sum_{j=0}^{5} I_j \sin{\left(4\pi \frac{2.5 - j}{6}\right)}. \label{eq-Fourier2}
\end{eqnarray}
In our implementation the filtergram intensities were calculated as
the convolution of the synthetic iron line with each of the six
transmission filter profiles.
The line amplitude and depth and the continuum intensity are then calculated 
according to the formulas:

\begin{equation} \label{eq-sigma}
	\sigma = \frac{T}{\pi \sqrt{6}}\sqrt{\ln{\left(\frac{a_1^2 + b_1^2}{a_2^2 + b_2^2}\right)}},
\end{equation}
\begin{equation} \label{eq-linedepth}
	I_d = \frac{T}{2\sigma \sqrt{\pi}}\sqrt{a_1^2 + b_1^2}\exp{\left(\frac{\pi^2 \sigma^2}{T^2}\right)},
\end{equation}
\begin{equation} \label{eq-continuum}
	I_c = \frac{1}{6} \sum_{j=0}^{5} I_j + I_d \exp{\left[-\frac{(\lambda_j - \lambda_0)^2}{\sigma^2}\right]}.
\end{equation} 
Here $T=412.8$ m{\AA} is six times the nominal wavelength separation between the filter transmission profiles
and $\lambda_j$ corresponds to the central wavelength of the $j$-th filter. 

The actual implementation of the MDI-like algorithm is 
slightly different than as explained above. Tests performed on Gaussian profiles have shown that
the line parameters calculated using the formulas above are affected
by inaccuracies generated by the finite term Fourier expansion (using
only six wavelengths to calculate the Fourier transform), the finite sampling of the line
and the
fact that the transmission profiles are not $\delta$-functions.
In particular, \cite{couvidat2012} reported that Equation~(\ref{eq-sigma}) overestimates the real  width by $\sim 20$\%
and that Equation~(\ref{eq-linedepth}) underestimates the actual  depth by $\sim 33$\%. To take into account 
these uncertainties, in the actual implementation Equation~(\ref{eq-sigma}) is multiplied by a factor of $5/6$
and Equation~(\ref{eq-linedepth}) is multiplied by a factor of  $6/5$. Tests have also revealed that the error in the continuum intensity is only $\sim 1\%$ so that
no correction factor is applied to Equation~(\ref{eq-continuum}).

Moreover, Equations (\ref{eq-linedepth}) and (\ref{eq-continuum}) are not computed using the $\sigma$ values
returned by Equation (\ref{eq-sigma}). The reason is that, as shown in previous sections,  the  $\sigma$ returned by Equation (\ref{eq-sigma})
suffers from uncertainties 
induced by strong deviations of the line shape from a Gaussian profile (like in presence of strong fields), which would cause uncertainties in the estimate 
of the other two line parameters. Instead, 
the algorithm uses standard values of $\sigma$ derived from a HMI
full-disk  line-width map observed in a day of low activity.
After the line width  is converted to $\sigma$, the standard $\sigma$ values are computed as the 
fifth order polynomial fit to the variation of the azimuthally averaged $\sigma$-values with radial distance from disk center.

\begin{figure}    
   \centerline{\includegraphics[width=8cm]{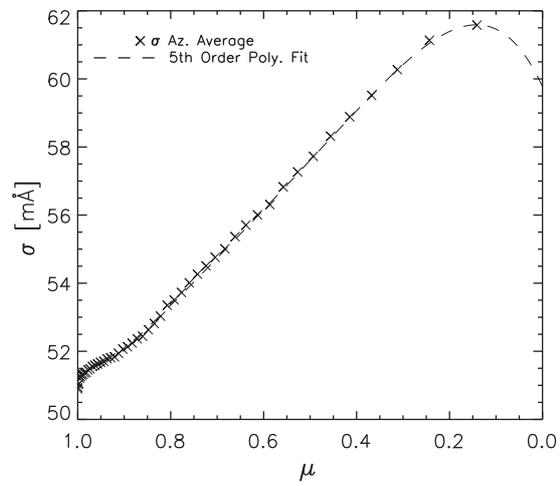}
              }
   \caption{Line-of-sight dependence of the $\sigma$ values employed
     to compute the line depth and the continuum intensity. The
     $x$-axis is the viewing angle $\mu = \cos{\theta}$ and the
     dashed line is a fifth-order polynomial fit to the data.}
    \label{Sigma}
  \end{figure}

The values of $\sigma$ as function of the cosine of the heliocentric angle
employed in our implementation are illustrated in Figure \ref{Sigma} and were obtained from a HMI line-width map taken on 
September 10, 2009.

\end{article} 
\end{document}